\documentclass{epl}
\title{Non-Conventional Structural Phase Transitions \\
and Amphiphobic Matter}

\shorttitle{Non-Conventional Structural Transitions}

\author{I. Erukhimovich}
\institute{Moscow State University, Moscow 119899 Russia}


\begin{document}

\maketitle

\begin{abstract}
The order-disorder and order-order transitions in the ternary
$ABC$ block and graft copolymers are analyzed via a Leibler-like
microscopic approach. We show that simple cubic, face-centered
cubic, well known double gyroid as well as some other phases could
be stable in these systems in a vicinity of the critical point,
equally with the conventional phases (body-centred cubic,
hexagonal planar and lamellar). In particular, the ternary linear
$ABC$ block copolymers with a long middle block non-selective with
respect to the side blocks are especially inclined to form the
gyroid phase. A new cubic non-centrosymmetric phase and some other
cubic phases are also first predicted to exist as the most stable
low temperature phase instead of the lamellar one. Such a phase
behavior is suggested to be common for a new class of materials we
call amphiphobic since their (macro)molecules consist al least of
three mutually incompatible types of monomers.
\end{abstract}

A number of systems undergo order-disorder and order-order phase
transitions due to the fact that their uniform state becomes
unstable with respect to certain spatial fluctuations of the
corresponding order parameter $\Phi (\bf{r})$ having a finite
period $L$ and, respectively, wave number $q_{*}=2\pi /L$. For
instance, addition of an ionic solute to a solvent in its critical
region may result in charge-density waves generation \cite
{Nabutov,Stell}, similar behavior was predicted and observed in
weakly charged polyelectrolytes
\cite{Borue88,JL90,Schlosseler91M,Shibayama92}. Microphase
separation in solutions and melts of copolymers is also due to
instability with respect to spatial fluctuations of the polymer
concentration having a finite period
\cite{deGennes79m,Leibler,Erukh82}. The various morphologies
emerging as a result of these transitions have attracted much
interest \cite{BF99} due to both numerous possible technological
applications and interesting physics underlying their formation.

A common theoretical framework for these systems is provided by
the weak crystallization theory
\cite{Leibler,Landau37,AlexMc,KLMreview}. As consistent with the
mean field approximation of this theory \cite{Leibler}, the
typical succession of the 1st order phase transitions occurring
with decrease of temperature is as follows: the uniform
(disordered) phase (DIS) - body-centred cubic lattice (BCC) -
hexagonal planar lattice (HEX) - lamellar structure (L). Further
we refer to the phases BCC, HEX and LAM as the conventional ones.
For some special sets of parameters, all these phase transitions
merge in the critical point where the 2nd order phase transition
from the disordered to lamellar phase occurs. For
$A_{fN}B_{(1-f)N}$ diblock copolymers such a parameter is just the
composition $f$ and the critical point corresponds to the
symmetric diblock copolymer ($f_{c}=0.5$) in case the repeated
units of both blocks have the same excluded volumes $v$ and Kuhn
lengths $a$. Other phases usually could exist as metastable only,
even though simple cubic (SC) and face-centered cubic (FCC) were
shown to become stable for some special models \cite{BDM}.

One more equilibrium phase commonly encountered in lipid-water and
surfactant systems \cite{Gyr0} has been observed in weakly
segregated molten diblock copolymers \cite{Gyr1,Gyr2}. It is
characterized by $Ia\overline{3}d\;$space group symmetry and often
called the double gyroid (G). The bicontinuous morphology
characteristic of this phase has attracted much interest during
the last decade and a few of theoretical papers has been published
to explain and describe this phase. In particular, for the
aforementioned copolymer systems the mean-field phase diagrams
were first theoretically calculated in the works \cite{MS}. A
specific feature of these
phase diagrams is the existence of two triple points ($f_1^{tr}<f_c$ and $%
f_2^{tr}>f_c$) in which three phases HEX, G and LAM coexist.
Therewith, the conventional sequence DIS-BCC-HEX-LAM and
non-conventional one DIS-BCC-HEX-G-LAM would hold for compositions
within and out of the interval, respectively. (For simplicity we
do not address in this paper the fluctuation-caused changes of the
mean-field phase diagrams discussed in detail in
\cite{Braz,BDM,FH87,BF91,Monica91,DEr91}.)

A general analysis carried out by the author \cite{me96} within
the mean-field approximation of the weak crystallization theory
led to conclusion that the existence of triple point(s) different
from the critical one is not a necessary feature of the phase
diagrams of the systems capable of forming the thermodynamically
stable bicontinuous gyroid morphology. The phase G as well as
other phases different from the conventional ones were predicted
to be stable, at certain conditions, around the critical point.
However, this behavior was not reported, basing on microscopic
calculations, for any real systems yet.

In this Letter we demonstrate that the real systems revealing the
non-conventional morphologies G, SC, FCC and defined below BCC$_2$
and G$_2$ in the close vicinity of the critical point in
accordance with our prediction \cite{me96} do exist. In
particular, we show that the systems especially inclined to form
the G phase are the ternary ABC block copolymers with a long
non-selective middle block.

We start with a general phenomenological analysis. Being
interested in the phase behavior close to the critical point, we
write down the free energy related to emergence of a non-uniform
scalar order parameter profile $\Phi ({\bf r})$ a Landau expansion
in powers of $\Phi $ up to the 4th order:
\begin{eqnarray}
\Delta F &=&\int \frac{\tau +\Gamma _{2}(q)-\Gamma _{2}(q_{\ast })}{2}\frac{%
\left| \Phi _{{\bf q}}\right| ^{2}d{\bf q}}{(2\pi )^{3}}{\bf
\;}+\Delta
F_{3}+\Delta F_{4}  \nonumber \\
\frac{\Delta F_{n}}{(2\pi )^{3}} &=&\frac{1}{n!}\int \delta \left(
\sum_{i=1}^{i=n}{\bf q}_{i}\right) \Gamma _{n}({\bf q}_{1}{\bf ...q}%
_{n})\prod_{l=1}^{l=n}\Phi ({\bf q}_{l})\frac{d{\bf q}_{l}{\bf
\;}}{(2\pi )^{3}}  \label{free}
\end{eqnarray}
The coefficients $\Gamma _{i}$ appearing in the free energy
expression (\ref {free}), depend on the structure of the system,
the function $\Gamma _{2}(q)$ has a minimum at $q=q_{\ast }$ and
$\tau $ is an effective dimensionless temperature measured from
the instability point.

The system morphology is described by the order parameter $\Phi
({\bf r})$ (or its Fourier transform $\Phi _{{\bf q}}$) providing
the minimum of this free energy.

In general, it can be expanded in an infinite series in the
Fourier harmonics corresponding to the set $\left\{ {\bf
n}_i\right\} $ of the points of the inverse lattice conjugated to
the chosen spatial lattice. But in the weak segregation
approximation (close to the critical point) one takes account only
of the $2k$ main harmonics belonging to the 1st coordination
sphere of the inverse lattice
\begin{equation}
\Phi _0({\bf r)}=A%
\mathop{\displaystyle \sum }
_{\left| {\bf n}_i\right| =1}\left( \exp i\left( q_{*}{\bf
n}_i{\bf r+}\phi _i\right) +c.c.\right) \;  \label{sum}
\end{equation}
Therewith, the phases $\phi _i$ are relevant. Indeed, substituting
the trial functions (\ref{sum}) into the r.h.s. of eq.~(\ref
{free}) and minimizing the result with respect to $A${\bf \ }gives
\begin{equation}
\Delta F=\tau A_0^2+\alpha _kA_0^3+\beta _kA_0^4\;\;\;
\label{fre}
\end{equation}
where $A_0=Ak^{1/2},\;\alpha _k=\left( \gamma /k^{3/2}\right)
\sum_3\cos \Omega _j^{(3)}$ and
\begin{equation}
\beta _k=\frac{\lambda _0(0)}{4k}+\frac{\sum \lambda
_0(h)+\sum_4\lambda (h_1,h_2,h_3)\cos \Omega _j^{(4)}}{k^2}
\label{coef}
\end{equation}
We use the Leibler designations and parameters \cite{Leibler}
\[
h_{1}=\left. \left( \mathbf{q}_{1}+\mathbf{q}_{2}\right)
^{2}\right/ q_{\ast }^{2},\;h_{2}=\left. \left(
\mathbf{q}_{1}+\mathbf{q}_{3}\right) ^{2}\right/ q_{\ast
}^{2},\;h_{3}=\left. \left( \mathbf{q}_{1}+\mathbf{q}_{4}\right)
^{2}\right/ q_{\ast }^{2}.\;
\]
$\gamma =$ $\Gamma _3({\bf p}_1,{\bf p}_2{\bf ,p}_3)\;\left( \left| {\bf p}%
_i\right| =q_{*}\right) ,$ $\lambda _0(h)=\lambda (0,h,4-h),$
$\lambda
(h_1,h_2,h_3)=\Gamma _4({\bf q}_1,{\bf q}_2,{\bf q}_3,{\bf q}_4)$ with $%
\sum_{i=1}^{i=4}{\bf q}_i=0,$ $\left| {\bf q}_i\right|
=q_{*}\;$\cite{comm}. The phases $\Omega _j^{(n)}$ are the
algebraic sums of the phases $\phi $ for the triplets and
noncoplanar quartets of the vectors involved in the definition of
corresponding $\gamma $ and $\lambda $, the symbol $\sum_n$
designating summation over all sets of such $n$ vectors. The first
summation
in eq.~(\ref{coef}) is over all pairs of noncollinear vectors ${\bf q}_i$ and $%
{\bf q}_j$.

{\bf The BCC family}. The six main harmonics
\begin{eqnarray}
\mathbf{n}_{1} &\sim &(0,1-1),\;\mathbf{n}_{2}\sim (-1,0,1),\;\mathbf{n}%
_{3}\sim (1,-1,0),\quad   \nonumber \\
\mathbf{n}_{I} &\sim &(0,-1-1),\;\mathbf{n}_{II}\sim
(-1,0,-1),\;\mathbf{n}_{III}\sim (-1,-1,0),\quad \label{nbcc}
\end{eqnarray}%
are known to correspond the conventional BCC morphology if all the
phases are zero. This set gives also the BCC$_2$ lattice
\cite{BDM} if we choose
\begin{equation}
\phi_{I}=\phi_{II}=\phi _{III}=0,\quad \phi_1=\phi_2=\phi_3=\pi /2
\label{Phibcc2}
\end{equation}
As consistent with the definitions (\ref{nbcc}), (\ref{Phibcc2}),
the order parameter (\ref{sum}) for BCC$_2$ reads
\[
\Phi \left( \mathbf{r}\right) =A\left( \cos \left( \widetilde{x}+\widetilde{y%
}\right) +\cos \left( \widetilde{y}+\widetilde{z}\right) +\cos
\left(
\widetilde{z}+\widetilde{x}\right) -\sin \left( \widetilde{x}-\widetilde{y}%
\right) -\sin \left( \widetilde{y}-\widetilde{z}\right) -\sin
\left( \widetilde{z}-\widetilde{x}\right) \right) ,
\]%
where the waved coordinates are scaled as compared to the original
ones according to the rule $\widetilde{s} = s q_{\ast }/\sqrt{2}$.
As consistent with the most general symmetry properties \cite{IT},
BCC$_2$ morphology is non-centrosymmetric. Remarkably, the zero
level surface for BCC$_2$ differs from the well known "gyroid"
surface \cite{MSRI} only in a shift of the origin of the
co-ordinate system.

{\bf The G family. }The 12 main harmonics
\begin{eqnarray}
\mathbf{n}_{01} &\sim &(-2,1,1),\;\mathbf{n}_{11}\sim (-2,-1-1),\;\mathbf{n}%
_{21}\sim (2,1,-1),\;\mathbf{n}_{31}\sim (2,-1,1),\;  \nonumber \\
\mathbf{n}_{02} &\sim &(1,-2,1),\;\mathbf{n}_{12}\sim (1,2,-1),\;\mathbf{n}%
_{22}\sim (-1,-2,-1),\;\mathbf{n}_{32}\sim (-1,2,1),  \nonumber \\
\mathbf{n}_{03} &\sim &(1,1,-2),\;\mathbf{n}_{13}\sim (1,-1,2),\;\mathbf{n}%
_{23}\sim (-1,1,2),\;\mathbf{n}_{33}\sim (-1,-1,-2),  \label{ngyr}
\end{eqnarray}
provide $i$) the bi-continuous gyroid morphology $Ia\overline{3}d$
with the phases
\begin{eqnarray}
\alpha _{12} &=&\alpha _{23}=\alpha _{31}=\alpha _{01}=\alpha
_{02}=\alpha
_{03}=0,  \label{Phig} \\
\alpha _{21} &=&\alpha _{32}=\alpha _{13}=\alpha _{11}=\alpha
_{22}=\alpha _{33}=\pi ;  \nonumber
\end{eqnarray}
$ii$) the morphology we call the BCC$_3$ if all the 12 phases
equal zero (it is just the ordinary BCC but the dominant harmonics
correspond to the 3rd rather than 1st co-ordination sphere); and
$iii$) the morphology we call the G$_2$ if $\alpha _{21}=\alpha
_{32}=\alpha _{13}=\pi $ and other 9 phases
equal zero (it seems to correspond to the crystallographic symmetry class $I%
\overline{4}3d$).

The next step is to take into account the explicit angle
dependence of the fourth vertex $\Gamma _4$ appearing in the
expression (\ref {free}) rather than to adopt the commonly
accepted approximation \cite{FH87}
\begin{equation}
\Gamma _4({\bf q}_1,{\bf q}_2,{\bf q}_3,{\bf q}_4)\approx \lambda
_0(0).
\end{equation}
Assuming the angle dependence to be given by the first
non-constant term in the expansion of $\Gamma _4$ in powers of
$h_i$ :
\begin{equation}
\Gamma _4(h_1,h_2,h_3)=\lambda _0\left( 1-\frac{3\delta
}{32}\left( 4^2-\sum_{i=1}^3h_i^2\right) \right)  \label{gam0}
\end{equation}
we build the phase diagram shown in fig.~\ref{fig1} on the plane
($\delta $ - the reduced temperature $\widetilde{\tau }=32\tau
\lambda _0/\left( 9\gamma ^2\right) $), where only the phase
transition lines starting at the
very critical point are shown. As seen from \ref{fig1}, for $\delta >\delta _0$ $%
=0.362$ the nonconventional sequences occur:

\begin{figure}
  \onefigure[width=14cm,height=5cm]{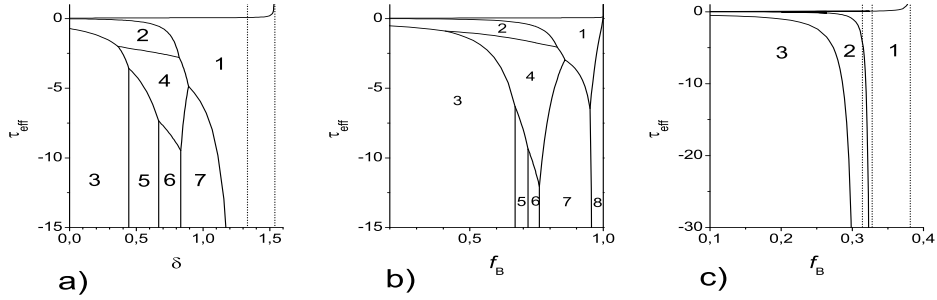}
  \caption{The phase diagrams of the ordered
phases whose existence is possible close to the critical point of
the order-disorder transitions. a) for the phenomenological angle
dependence \ref{gam0} of the forth vertex on the plane (effective
temperature $\tau _{eff}$ - structure parameter $\delta $). b) for
the ternary $ABC$ linear triblock copolymers with the middle
non-selective block on the plane ($\tau _{eff}$ - $f_B$); c) for
the $ABC$ miktoarm terpolymers with one non-selective block
(refered to as $B$) on the plane ($\tau _{eff}$ - $f_B$).
The phases are designated as follows: {\it 1 }%
- BCC, {\it 2}{\em \ }-{\em \ }HEX, {\it 3 }- LAM, {\it 4 }- G,
{\it 5 }- BCC$_2$, {\it 6 }- FCC, {\it 7 }- SC, {\it 8 }- G$_2.$}
 \label{fig1}
 \end{figure}

DIS-BCC-HEX-G-LAM for $\delta _{12} > \delta > \delta _0$, $\delta
_{12}$ = $4/9$; \

DIS-BCC-HEX-G-BCC$_2$ for $\delta _{23} > \delta >
\delta_{12},\delta _{23}$ = $2/3$;

DIS-BCC-HEX-G-FCC for $\delta_1 > \delta > \delta_{23},
\delta_1$=$0.822$;

DIS-BCC-G-FCC for $\delta _{34} > \delta > \delta_1,
\delta_{34}$=$5/6$;

DIS-BCC-G-SC for $\delta _{45} > \delta > \delta _{34},
\delta_{45}$=$0.891$;

DIS-BCC-SC for $\delta _2 > \delta > \delta _{45},
\delta_2$=$4/3$;

DIS-BCC for $\delta _3 > \delta > \delta _2, \delta _3$=$1.538$.

The lines $\delta_2$=$4/3$ and $\delta_3$=$1.538$ are shown in
fig.~\ref{fig1} by the dotted lines. At last, for $\delta
> \delta _3$ the value of $\beta _{BCC}$ becomes negative and the
weak crystallization theory does not hold anymore even in the
vicinity of the critical point. In this region the proper
consideration of the ordered phases is possible with due regard
for the higher terms of the Landau expansion (\ref{free}) only.

Thus, to find any non-conventional phase transition sequence it is
intrinsically necessary to take into account the angle dependence.

Let us remember now that the phase diagrams calculated for the
diblock and star AB block copolymers \cite{Leibler,DE93} are quite
conventional in the sense that the gyroid phase exists only
outside an interval including the critical point \cite{MS}. We
conclude, therefore, that $\delta >\delta _0 $ for these systems
and the gyroid phase stability here is due to the ''higher
harmonics'' contribution \cite{Floudas,Vita,OM}. The same topology
is characteristic of the phase diagrams we calculated for some
other $AB$ molten copolymers of complex architecture.

The situation changes drastically if we address the ternary $ABC$
copolymers. We analyze them via numerical calculation of the
explicit expressions for the vertices $\Gamma _i$, their
substitution into (\ref{free}) and minimization of the obtained
free energy expressions with respect to the trial order parameters
(\ref{sum}) for all the morphologies described above. Therewith,
we employ an approximate reduction of the real 2-order parameter
problem to an effective 1-order parameter one. To this end, we
introduce a quantitative distinction between strongly and weakly
fluctuating order parameters $\Phi ({\bf r})$ and $\Psi ({\bf
r})$, respectively, minimize the free energy with respect to the
weakly fluctuating field $\Psi ({\bf r})$ given a profile of $\Phi
({\bf r})$ and, by doing so, obtain an effective
1-order parameter Hamiltonian in terms of the strongly fluctuating field $%
\Phi ({\bf r})$. The whole procedure is described in detail in
refs \cite{Floudas,Vita,OM,FL89,me94,abc97}.

The last simplification is the approximation of the binary
interaction parameters via the solubility parameters: $2\chi
_{ij}=v\left( \delta _i-\delta _j\right) ^2/\left( 2T\right) $,
where the temperature $T$ is measured in energetic units,
$\delta_i$ is the conventional
(temperature-independent) solubility parameter of the $i$-th component and $%
v $ is (the same) excluded volume of the repeating units $A,B,C$.
Then there are two independent interaction parameters naturally
characterizing the ternary systems:

\begin{equation}
\chi _{AC}=\frac{v\left( \delta _A-\delta _C\right) ^2}{2T},\qquad x=\frac{%
2\delta _B-\delta _A-\delta _C}{\delta _A-\delta _C}
\end{equation}

The first of them characterizes incompatibility of the side blocks in the $%
ABC$ triblock copolymer whereas the selectivity parameter $x$
describes how much is the middle block $B$ selective with respect
to the side blocks \cite {abc97}.

It could be shown via considerations similar to those presented in
\cite{abc97} that for $x=0$ the cubic term vanishes identically along the line $%
f_A=f_C$ both for molten $A_lB_mC_n$ triblock and trigraft
copolymers. Thus, this line is the critical one for these systems.
Their phase diagrams in the vicinity of this critical line we
calculated using the procedure and
approximations described above are presented in fig.~\ref{fig1} b, c on the plane ($f_B$%
, the reduced temperature $\widetilde{\tau }=32\tau \lambda
_0/\left( 9\gamma ^2(f_B)\sigma ^2\right) $), where the asymmetry
parameter $\sigma =\left| l-n\right| /(l+n)$ and a scaling factor
$\gamma (f_B)$ are introduced.

\begin{figure}
  \oneimage[width=10cm,height=5cm]{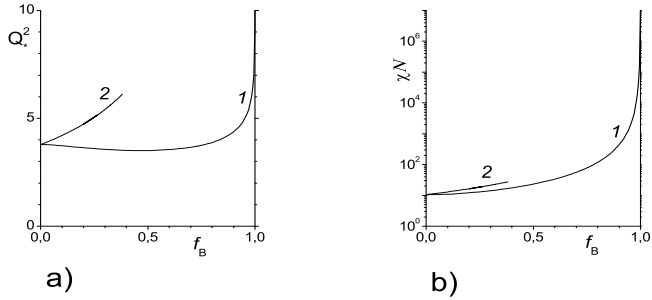}
  \caption{The characteristic scale $Q_{*}^2=\left( q_{*}a\right) ^2N/6$ (a) and location of the
spinodal with respect to microphase separation
$\widetilde{\chi}=\chi$ $N$ (b) as functions of the non-selective
block composition $f_B$ for the linear ($1$) and star ($2$)
ternary block copolymers.}
 \label{fig2}
 \end{figure}

Whereas $\widetilde{\tau }$ describes the distance between the
reduced temperatures of different phase transition lines and that
of spinodal, location of the spinodal itself is naturally
described in terms of the reduced parameter $\widetilde{\chi
}=\chi _{AC}N$, where $N=l+m+n$ is the
total degree of polymerization of the $ABC$ macromolecule. The values of $%
\widetilde{\chi }$ and the reduced value of the critical wave number $%
Q_{*}=\left( q_{*}a\sqrt{N}\right) /6$ are plotted as functions of
the middle block composition $f_B$ on fig.~\ref{fig2}.
We see clearly that the evolution of the phase diagrams of the
linear $ABC$ block copolymers with increase of the middle block
fraction $f_B$ closely follows the trend predicted
phenomenologically in \cite{me96} and shown in fig.~\ref{fig1}a.
Namely, a consecutive replacing of the lamellar morphology as the
most low-temperature stable phase according to sequence LAM-BCC$_2$-FCC-SC-G$%
_2$ occurs with decrease of the length of the shortest side block.
This trend is quite natural and means that it is impossible to
form lamellae when the side blocks are short enough. The only
topological difference between the phase diagrams shown on
fig.~\ref{fig1}a and fig.~\ref{fig1}b is replacing of the BCC
phase by a new phase of symmetry G$_2$ for extremely short (but
finite) side blocks with $f_{B}>0.96$. This difference is,
obviously, due to the fact that the angle dependence of the vertex
$\Gamma _4$ cannot be approximated by the simple parabolic form
\ref{gam0} if it is strong enough. Some parallels with our
predicitons could be found in \cite{Mogi94,Matsen}.

Even more strong and different is this dependence for the molten
$ABC$ miktoarm terpolymers with one arm (refered to as $B$)
non-selective with respect to both other as demonstrated by the
topology of their phase diagram shown on fig.~\ref{fig1}c. The
weak segregation theory is intrinsically not applicable for these
systems for $f_B>0.3815$ since the forth vertex of the BCC phase
becomes negative here and, therefore, this phase becomes unstable
with respect to strong segregation. In this case the sequence of
the most low-temperature stable phases is as follows: LAM-HEX-BCC.
A rich phase behavior was found also in our preliminary analysis
of the mixtures of ternary and binary block copolymers, the values
of $f_B$ corresponding to the phase transitions
LAM-BCC$_2$-FCC-SC-G$_2$ being strongly dependent on the
concentration and the polymer length ratio.

Summarizing, in this paper we carried out a microscopic
Leibler-like analysis of the ternary triblock copolymers with the
middle block non-selective with respect to the side ones. We
showed that for the linear ternary triblock copolymers \emph{i}) a
new cubic non-centrosymmetric morphology BCC$_2$ should replace
the lamellar one as the most stable low temperature phase for
reasonably long middle block ($0.67<f_{B}<0.72$), with further
increase of $f_B$ BCC$_2$, in turn, will be replaced by SC, FCC
and G$_2$ morphologies; \emph{ii}) the gyroid phase
$Ia\overline{3}d$ becomes stable in the very vicinity of the
critical point for $0.38<f_{B}<0.85$. On the contary, for the
ternary miktoarm (star) triblock copolymers we predict increase of
the BCC stability and strong segregation with increase of $f_B$.
One can expect, in general, that the phase behavior of block
copolymers with $n\geq 3$ mutually incompatible sorts of blocks
would be not only much richer but also much more architecture
dependent than that of the conventional binary block copolymers.
We believe this class of materials to deserve a particular generic
name of {\it amphiphobic matter}.

\acknowledgments I greatly acknowledge the support by INTAS and
DFG as well as many valuable discussions with V. Abetz at the
initial part of this work and the Humboldt Foundation support and
hospitality of K. Binder I benefited from when finalizing the
paper.


\begin{thebibliography}{0}
\bibitem{comm}
The value of the 4-th vertex may depend also on the value of
$q_{*}$ (e.g., for block copolymers). However, it is only
$h$-(''angle'') dependence that affects the relative stability of
the different phases in the mean field apptoximation.
\end{thebibliography}
\end{document}